\documentstyle[12pt,aasms4]{article}

\slugcomment{Accepted to be published in The Astrophysical Journal Supplement Series}

\lefthead{Gonz\'alez Delgado \& Leitherer}
\righthead{Stellar Library}

\begin{document}

\title{Synthetic spectra of H Balmer and HeI absorption lines. I: Stellar Library}

\author{Rosa M. Gonz\'alez Delgado}
\affil{Instituto de Astrof\'\i sica de Andaluc\'\i a, Apdo. 3004, 18080 Granada, Spain}
\affil{Electronic mail: rosa@iaa.csic.es}

 \and 

\author{Claus Leitherer}
\affil{Space Telescope Science Institute, 3700 San Martin Drive, Baltimore, MD
21218}
\affil{Electronic mail: leitherer@stsci.edu}



\newpage

\begin{abstract}

We present a grid of synthetic profiles of stellar H Balmer and HeI lines  at
optical wavelengths with a sampling of 0.3 \AA. The grid spans a range of 
effective temperature 50000 K $\geq$ T$_{eff}$ $\geq$ 4000 K, and  gravity 
0.0$\leq$log$ g\leq$5.0 at solar metallicity. For $T_{eff}\geq$25000 K,
NLTE  stellar atmosphere models are computed using the code TLUSTY (Hubeny
1988). For cooler  stars, Kurucz (1993) LTE models are used to compute the
synthetic spectra. The grid includes the profiles of the high-order  hydrogen
Balmer series and HeI lines for effective temperatures and gravities that  have
not been previously synthesized.  The behavior of H8 to H13 and HeI
$\lambda$3819 with effective temperature and gravity is very similar to that of
the lower terms of the series (e.g. H$\beta$) and  the other HeI lines at longer
wavelengths; therefore, they are suited for the determination  of the
atmospheric parameters of stars. These lines are potentially important to make
predictions for these stellar absorption features in galaxies with active star
formation. Evolutionary synthesis models of these lines for starburst and
post-starburst galaxies are presented in a companion paper. The full set of the
synthetic stellar spectra is available for retrieval at our website
http://www.iaa.es/ae/e2.html and http://www.stsci.edu/science/starburst/ or on
request from the authors at rosa@iaa.es.  

\end{abstract}



\newpage

\section{Introduction}

The dominant characteristics of the optical spectrum of starburst galaxies are
its nebular emission lines. These lines are formed in the surrounding
interstellar medium of the starburst that is photoionized by photons emitted by
stars with mass higher than 10 M$\odot$. These massive stars dominate the
ultraviolet and contribute significantly to the optical continuum, that comes
mainly from A, B and O stars. The most conspicuous features of the spectra of
these early stars are their strong hydrogen Balmer and neutral helium absorption
lines. In the spectrum of a starburst, the nebular emission lines are
superimposed on the stellar absorption lines, and usually emission dominates
over absorption. However, the contribution of the underlying absorption becomes
increasingly important for higher terms of the Balmer series (H$\gamma$,
H$\delta$,...) and some of the HeI lines ($\lambda$ 4471, 4387, 4026, ...). Very
often, the effect of the stellar population is also seen as absorption wings in
the H$\beta$, H$\gamma$ and H$\delta$ lines. The observation of these lines with
high spectral resolution makes it possible to estimate the contribution of the
underlying absorption. On the other hand, the analysis of the profiles
high-order of the Balmer series, which are dominated by absorption instead of
emission, can yield information on the properties of the stellar content of the
starbursts and on their evolutionary stage. Therefore, high-resolution
absorption profiles of the high-term Balmer and He I lines covering a wide range
in effective temperature and gravity are needed to predict the composite stellar
spectra of starburst galaxies. 

Observational and synthetic stellar spectra are available in the literature.
However, the observational atlases have only intermediate or low spectral
resolution (Burstein et al. 1984; Jacoby, Hunter \& Christian 1984; Walborn \& Fitzpatrick 1990;  Cananzi, Augarde, \& Lequeux 1993) and the synthetic stellar
libraries do not cover the high-order Balmer series and HeI lines (Auer \&
Mihalas 1972; Kurucz 1979, 1993). For these reasons, a grid of stellar
atmospheres and synthetic spectra in the wavelength range from 3700 to 5000 \AA\
with a sampling of 0.3 \AA\ and covering 50000 to 4000 K has been computed using
the code of Hubeny (1988). In this paper, we present the grid of synthetic
profiles, which are compared with observations and synthetic profiles computed
by Auer \& Mihalas (1972) and Kurucz (1993). In a companion paper (Gonz\'alez
Delgado, Leitherer \& Heckman 1999; hereafter paper II), evolutionary synthesis
profiles of H and He absorption lines for star forming regions are presented. 
 
\section{Stellar library}

\subsection{The Grid}

The grid includes the synthetic stellar spectra of the most relevant H and He I
lines from 3700 to 5000 \AA, in five different wavelength ranges (Table 1).
H$\epsilon$ is not synthesized because it is coincident with the CaII H line,
and this line shows a very strong interstellar component in individual stars
and in the integrated spectrum of a galaxy. HeII lines, He I $\lambda$5876, HeI $\lambda$6678 and H$\alpha$ are not synthesized because in hot stars the
profiles of these lines are affected by stellar winds (Gabler et al. 1989;
Herrero et al. 1992; Bianchi et al. 1994). However, HeI at wavelengths shorter
than 5000 \AA, H$\beta$ and the higher order-terms of the Balmer series are
only partially (mainly the core of the line but not the wings) or not at all
affected by sphericity and winds (Gabler et al. 1989). The spectra span a range
of effective temperature from 4000 to 50000 K, with a variable step from 500 K
to 5000 K, and a surface gravity log$g$=0.0 to 5.0 with a step of 0.5 (Table
2). The metallicity is solar. 

The spectra are generated in three different stages with a set of computer
programs developed by Hubeny and collaborators (Hubeny 1988; Hubeny \& Lanz
1995a; Hubeny, Lanz, \& Jeffery 1995). First, the stellar atmosphere is
calculated; then, the stellar spectrum is synthesized, and finally the
instrumental and rotational convolution are performed. 

\subsection{The model atmospheres of the grid}

 To generate a synthetic stellar spectrum, a model atmosphere is needed. For
$T_{eff}\geq$25000 K, the atmosphere is produced using version 193 of the
program TLUSTY (Hubeny 1988; Hubeny \& Lanz 1995a,b); for $T_{eff}\leq$25000 K
we use a Kurucz (1993) LTE atmosphere. TLUSTY calculates a plane-parallel,
horizontally homogeneous model stellar atmosphere in radiative and hydrostatic
equilibrium. The program allows departures from local thermodynamic equilibrium
and metal line blanketing, using the hybrid complete linearization and
accelerated lambda iteration (CL/ALI) method. 

To reduce computational time, non-blanketed non-LTE (NLTE) models are computed
with TLUSTY. H and He are considered explicitly. The population of their levels
(9 atomic energy levels of HI, 14 levels of HeI and 14 levels of HeII are
considered) are determined by solving the corresponding statistical equilibrium
equations. 25 additional atoms and ions contribute to the total number of
particles and to the total charge, but not to the opacity. 

The NLTE models are computed in three stages. First, an LTE-gray atmosphere is
generated. Here, the opacity is independent of wavelength, and the populations
of the energy levels are calculated assuming the local value of the temperature
and electron density. This model is used as a starting approximation for the
LTE model. Finally, the NLTE model is computed, where the gas and the radiation
are coupled. Here, departures from LTE are allowed for 39 energy levels.
Convection is suppressed in all the models. A depth-independent turbulence
velocity of 2 km s$^{-1}$ is assumed. Doppler broadening is assumed for all the line
transitions. The properties of the atmosphere are calculated at 54 depth
points. 

We use Kurucz (1993) LTE atmospheres for $T_{eff}\leq$25000 K because for stars
cooler than B1, NLTE effects are not very important, These models are line
blanketed, and we take the models with a turbulence velocity of 2  km s$^{-1}$
and solar metallicity.

\subsection{The synthetic profiles}

The synthetic spectra are computed with the program SYNSPEC (Hubeny, Lanz, \&
Jeffery 1995). This program reads the input model atmosphere, either calculated
by TLUSTY or from the Kurucz models, and solves the radiative transfer
equation, wavelength by wavelength in a specified spectral range. The program
also uses an input line list that contains the transitions in the six specified
wavelength ranges that we synthesize here. The line list has a format similar
to the Kurucz \& Peytremann (1975) tables. 

The continuum opacity is calculated exactly the same way as it was done in the
atmosphere model. For $T_{eff}\geq$ 25000 K, only continuum opacities from the
atomic energy levels of H and He are considered. The opacity sources are: 1)
photoionizations from all the explicit levels (39 levels in TLUSTY models); 2)
free-free opacity for all the explicit ions (HI, HeI and HeII in TLUSTY models
and H, Mg, Al, Si and Fe in the Kurucz models); 3) electron scattering. The
line opacity is calculated from the line list. Although the structure of the
atmosphere (for $T_{eff}\geq$ 25000 K) is calculated with only H and He, the
synthetic spectra are calculated assuming line blanketing from elements with
atomic number Z$\leq$28. The atoms and transitions that were treated in NLTE in
the atmosphere models are also treated in NLTE by SYNSPEC. For those lines that
originate between levels for which the population is calculated in LTE, SYNSPEC
uses an approximate NLTE, based on the second-order escape probability theory
(Hubeny et al. 1986). 

The line profiles have the form of a Voigt function and take into account the
effect of natural, Stark, Van der Waals and thermal Doppler broadening. The
broadening parameters for H and He are those tabulated by Vidal, Cooper, \&
Smith (1973) for the first four members of the Balmer series, and up to H10 by
Butler (private communication). For HeI, the line broadening tables for
$\lambda$4471 are from Barnard, Cooper, \& Smith (1974) and those for 
$\lambda$4026, 4387, 4922 are from Shamey (1969). The maximum distance of two
neighboring frequency  points for evaluating the spectrum is 0.01 \AA. A
turbulence velocity of 2 km s$^{-1}$ is assumed. 

Finally, the program ROTIN performs the rotational and instrumental
convolution. A FWHM instrumental profile of 0.01 \AA\ and a rotational velocity
of 100 km s$^{-1}$ are assumed which is the typical rotational velocities  in
massive stars (Conti \& Ebbets, 1977). The final sampling of the spectra is 0.3
\AA.

\section{Results}

The synthetic spectra of the grid are available for retrieval at
http://www.iaa.es/ae/e2.html and http://www.stsci.edu/science/starburst/.
Figure 1 compares the spectra for $T_{eff}$=25000 K and log$ g$= 4.0 computed
with TLUSTY+SYNSPEC assuming NLTE with those computed with SYNSPEC and Kurucz
LTE atmospheres. The two spectra are essentially identical; this justifies the
use of LTE stellar atmosphere models for $T_{eff}\leq$25000 K. Figure 2 shows
the spectra of a typical O ($T_{eff}$=40000 K, log$ g$=4.0), B ($T_{eff}$=20000
K, log$ g$=4.0) and A ($T_{eff}$=10000 K, log$ g$=4.0) star. The most important
lines in the spectrum are labelled. The equivalent width of the most important
H and HeI lines have been measured in the synthetic spectra. Several
measurements were done with different spectral windows. This allows a
calibration of the contribution of weaker lines to the spectral index that
characterizes each of the Balmer lines. The spectral index was measured from
the continuum equal to 1 (this represents the real equivalent width of the
lines), but also from a pseudo-continuum that was determined by fitting a first
order polynomial (except for the spectral range 3700-3900 \AA, for which we
used a third order) to the windows defined in Table 3. This spectral index
simulates the equivalent width we can measure in observed stellar spectra.
Tables 4 to 7 show the equivalent widths of H$\delta$, H8, HeI $\lambda$4471
and HeI $\lambda$3819, respectively. 

Figure 3 shows the equivalent widths of H$\beta$, H$\delta$ and two of the
high-order terms of the Balmer series (H8 and H9) as a function of $T_{eff}$
for main sequence stars (log$ g$= 4.0). The maximum equivalent width occurs at
$\sim$9000 K,  corresponding to an early-type A star. The plot indicates that
the high-order Balmer series lines show, like the lower terms, a strong
dependence on $T_{eff}$ and on gravity (Figure 4); thus, they are an efficient
tool for the determination of the fundamental stellar parameters. Figure 5
shows the equivalent width of the HeI lines as a function of $T_{eff}$ for main
sequence stars (log$ g$= 4.0) (see also Tables 6 and 7 for HeI $\lambda$4471
and  HeI $\lambda$3819, respectively). They also show a strong dependence on
$T_{eff}$ and gravity (Figure 6). The maximum is at $\sim$20000 K, corresponding
to an early-type B star. The increase of the equivalent width at
$T_{eff}\leq$10000 K is not due to HeI absorption, but to the contribution of
some metal lines (FeII $\lambda$4385, FeI $\lambda$4920, FeII $\lambda$4924) 
which fall in the windows where the equivalent width is measured. 
 
\section{Test of the profiles}

In this section the synthetic profiles of some of the H and HeI lines are
compared to the  Auer \& Mihalas (1972) and  Kurucz (1993) profiles. They are
also compared with observations. The goal is to test the advantages and
limitations of our grid with respect to previous work.

\subsection{Comparison to Auer-Mihalas models}

Auer \& Mihalas (1972) computed non-blanketed NLTE profiles of H (P$\alpha$,
H$\alpha$, H$\beta$, H$\gamma$ and L$\alpha$) and HeI ($\lambda$4026, 4387,
4471 and 4922) for  $T_{eff}$ from 30000 to 50000 K, and log$ g$=3.3 to 4.5.
Figure 7 compares the profiles of H$\gamma$ for $T_{eff}$=30000 and 40000 and
log$ g$=4.0 with the spectra of the grid. The profiles are very similar, the
small discrepancy comes from the effect of rotation (we assume a  rotation of 
100 km s$^{-1}$ and none in the Auer-Mihalas models) and the inclusion of
atomic transitions in  the spectral range we have synthesized. The contribution
of these lines makes the profile of H$\gamma$ asymmetric in our spectra but not
in the Auer-Mihalas profiles. This represents  an improvement of our synthetic
profiles over those of Auer \& Mihalas because our grid is more similar to the 
observed spectra of stars (see for example Walborn \& Fitzpatrick 1990).

Figure 8 shows the profile of HeI $\lambda$4471 for $T_{eff}$=40000 K and
log$ g$=4.0.  The apparent disagreement comes from the effect of the rotational
convolution performed in the grid.  In fact, the difference between the
profiles disappears when the comparison is done between  non-convolved
profiles. The effect of the rotational convolution is more drastic in the
profile of  the HeI lines than in the Balmer lines because HeI lines are weaker
and narrower than Balmer lines. 

\subsection{Comparison to Kurucz models.}

For $T_{eff}\leq$ 25000 K, the grid has been generated using the stellar 
atmosphere structure calculated by Kurucz (1993). Kurucz also synthesizes the 
profiles of H$\alpha$, H$\beta$, H$\gamma$  and H$\delta$. However, the
synthesis does not include the atomic transitions  that fall within  $\pm$100
\AA\ of the center of the Balmer lines. Figure 9 compares the profiles  of our
grid with those of Kurucz (1993). They are very similar, the apparent 
disagreement at $T_{eff}\leq$ 7000 K is due to the effect of the rotational 
convolution and the inclusion of metallic lines in the spectral ranges 
synthetized here. However, note that the profile of the Balmer lines before 
performing the rotation is equal to the profile synthesized by Kurucz.

Our synthetic spectra assume fewer sources of continuum and line opacity than 
the Kurucz stellar atmosphere. This inconsistency leads to differences between
the continuum  calculated by TLUSTY+SYNSPEC and the low resolution spectra
generated by Kurucz. However,  the difference is less than 8$\%$ in  the whole
spectral range synthesized here if $T_{eff}\geq$ 7000 K. For lower 
effective temperature, the deviation is more important since the shape of the
continuum also changes.  However, the normalized profiles of the Balmer lines
of our grid and Kurucz (1993)  are the same for all effective temperatures
($T_{eff}\leq$ 25000 K). Thus, we can confidently use the normalized 
profiles of the grid in our evolutionary synthesis code, where these profiles
are  calibrated in absolute flux.

\subsection{Comparison to observations}

Figure 10 compares the synthetic profiles from 3700 to 3900 \AA\ with the
spectra of observed stars from the stellar library of Jacoby, Hunter, \&
Christian (1984). The observed spectra are normalized assuming a
pseudo-continuum that was defined by fitting a third order polynomial to the
windows defined in Table 3. The synthetic spectra are binned to the resolution
of the observations ($\sim$4 \AA). There is good agreement between the
synthetic and observed profiles. The discrepancy between the observations of
the O5V star and synthetic profiles at  $\lambda\leq$3760 \AA\ is probably due
to uncertainties in the wavelength and flux calibration of the data. 

We have also compared the synthetic spectra with two stars observed by one of
us for a different project. The stars HD24760 and HD31295 are classified as
B0.5V and A0V, respectively. They were observed with the 2.5m Isaac Newton
Telescope at the Roque de los Muchachos Observatory using the 500 mm camera of
the Intermediate Dispersion Spectrograph and a TEK CCD detector. The dispersion
is 0.4 \AA/pix. Figures 11 and 12 compare the observed normalized spectra with
the synthetic profiles with effective temperature of 27500 K and 9500 K,
respectively. In both cases, the gravity is log$ g$=4.0. The agreement between
observations and synthetic spectra is very good, even if an optimization of the
fit was not attempted; we have only taken the spectra of our grid with values
of the $T_{eff}$ and gravity closer to the characteristic values of B0.5V and
A0V stars. 

\subsection{Effect of the metallicity}

Opacities play an important role in determing the properties of the structure
of the atmosphere. Balmer lines, however, are not very much affected by the
abundance of elements heavier than H and He if the temperature is higher than
7000 K. The comparison of the Balmer profiles synthesized by Kurucz (1993) for
metallicity between Z$\odot$ and Z$\odot$/10 shows that the profiles are
essentially the same in this range of metallicity if $T_{eff}\geq$7000 K. Thus,
we can use our normalized spectra to predict the synthetic spectra of a stellar
population younger than 1 Gyr if the metallicity is higher than Z$\odot$/10. 

\section{Summary}

We have computed a grid of stellar atmosphere models and synthetic spectra
covering five spectral ranges between 3700 and 5000 \AA\ that include the
profiles of the Balmer (H13, H12, H11, H10, H9, H8, H$\delta$, H$\gamma$ and
H$\beta$) and HeI ($\lambda$3819, $\lambda$4009, $\lambda$4026, $\lambda$4120, 
$\lambda$4144, $\lambda$4387, $\lambda$4471, and $\lambda$4922) lines with a
sampling of 0.3 \AA. The grid spans a range of effective temperature  50000
K$\geq T_{eff} \geq$4000 K, and gravity 0.0$\leq$log$ g\leq$5.0 at solar
metallicity. 

The spectra are generated using a set of computer programs developed by Hubeny
et al. (1995a,b). The profiles are generated in three stages. First, for 
$T_{eff}\geq$25000 K, we use the TLUSTY code (Hubeny 1988) to compute NLTE
stellar atmosphere models. We assume 9 energy levels of HI, 14 levels of HeI
and 14 levels of HeII explicitly in NLTE. For $T_{eff}\leq$25000 K, we take the
Kurucz (1993) LTE atmospheres. In the second stage, the synthetic spectra are
produced with the program SYNSPEC (Hubeny et al. 1995b) that solves the
radiative transfer equation using as input the model atmosphere and a line list
that contains information about atomic transitions in the relevant spectral
wavelength ranges. Although the NLTE models generated with TLUSTY are pure H
and He stellar atmospheres, SYNSPEC assumes line blanketing for elements
heavier than H and He. 

Our grid of synthetic spectra has limitations due to the inconsistencies
between the continuum and line opacities assumed in the stellar atmosphere and
those assumed in calculating the transfer of the lines, which start to be
important for $T_{eff}\leq$7000 K. However, they reproduce very accurate the
Balmer lines generated by Kurucz (1993) even at $\leq$7000 K. The profiles of
the HeI lines are also very similar to those generated by Auer \& Mihalas (1972)
for $T_{eff}$ between 30000 and 50000 K. 

This work was motivated by the importance of the HeI and the high-order Balmer
lines in the study of the stellar population of galaxies with active star
formation. These galaxies show Balmer and He lines in emission produced in
their HII regions that are super-imposed on the stellar absorption lines.
However, the equivalent width of the nebular lines decreases quickly with
decreasing wavelength, whereas the equivalent width of the stellar absorption
lines is constant with wavelength (e.g. the equivalent width of H8 ranges
between 2 and 12 \AA\ and H$\beta$ between 2.5 and 14 \AA). Thus, the
high-order Balmer series members are less contaminated by emission, and they
can be very useful to study the underlying stellar population in starburst
galaxies. The evolutionary synthesis profiles of H and He absorption lines for
starbursts and post-starbursts are presented in paper II. The full set of the
resulting models is available at our websites http://www.iaa.es/ae/e2.html and
http://www.stsci.edu/science/starburst/, or on request from the authors at
rosa@iaa.es.




{\bf Acknowledgments}

We thank Ivan Hubeny for his help during the initial phase of this project and
for making his code available to the community, and to Enrique P\'erez, Tim
Heckman and Angeles D\'\i az for their helpful suggestions and comments. The
spectra of some observed stars come from data of an ongoing project to build an
observational library of stellar spectra in collaboration with Angeles D\'\i az,
Pepe V\'\i lchez and Enrique P\'erez. This work was supported by the Spanish
DGICYT grant PB93-0139.

%
%

\clearpage

\figcaption{Synthetic spectra from 3700 to 42000 \AA\  for Teff= 25000
and log g=4.0 generated with TLUSTY, SYNSPEC and ROTIN assuming NLTE (full line)
and with Kurucz (1993) LTE stellar atmosphere models and  SYNSPEC and ROTIN (dotted
line). }

\figcaption{Synthetic spectra from 3700 to 5000 \AA\  generated with 
TLUSTY, SYNSPEC and ROTIN for: a) Teff=40000 and log g=4.0 (NLTE model). b)
Teff=20000 and log g=4.0 (LTE model). c) Teff=10000 and log g=4.0 (LTE model). A
rotational convolution of 100 km s$^{-1}$ is performed. The sampling of the spectra
is 0.3 \AA.}

\figcaption{Equivalent width of the Balmer lines as a function of the
effective  temperature for log g= 4.0. The equivalent width is measured in windows
of 60 \AA\   (for H$\beta$ and H$\delta$) and 30 \AA\ (for H8 and H9) width. The
flux is integrated  under a pseudo-continuum defined after fitting a polynomial in
the continuum windows (see Table 3).}

\figcaption{Equivalent width of H8 as a function of the effective
temperature. Each curve  is for a value of the gravity, from log g=5.0 to 1.5 in
steps of 0.5. }

\figcaption{Equivalent width of HeI ($\lambda$4026, 4388, 4471 and 4922)
lines as a  function of the effective temperature for log g= 4.5. The lines are
stronger in B stars;  and the maximum strength occurs at $\sim$ 20000 K. The
increase of the spectral index as the effective temperature decreases (for
Teff$\leq$ 10000 K) is due to the contribution of metal lines (for example FeII
$\lambda$4385,  FeI $\lambda$4920, FeII $\lambda$4924) to the spectral ranges used
to measure the equivalent width. The equivalent width of HeI lines for
Teff$\leq$10000 K is null.}

\figcaption{Equivalent width of HeI $\lambda$4026 as a function of the
effective temperature. Each curve is for a value of the gravity, from log g=5.0 to
2.5 in steps of 0.5. }

\figcaption{NLTE Auer \& Mihalas (1972) profiles of H$\gamma$ compared to
our NLTE synthetic  spectra for: a) Teff=30000 K , log g= 4.0. b) Teff=40000 K, log
g=4.0. The profiles are very  similar, the difference results from the rotation and
the  contribution of other atomic lines that fall in the wavelength range
synthetized in our grid but not in the Kurucz profiles.}

\figcaption{NLTE Auer \& Mihalas (1972) profiles of the line HeI
$\lambda$4471 for  Teff=40000 K and log g=4.0 compared with the synthetic spectra
of our grid.  The thin line is our synthetic profile without the effect of the
rotational  convolution and dashed line after performing the rotational convolution
of  100 km s$^{-1}$. The Auer \& Mihalas profile is shown as a dotted line. Note
that  the profiles are equal when no rotation is assumed. }

\figcaption{ LTE Kurucz (1993) profiles of the Balmer lines compared with
LTE  synthetic profiles of our grid for: a) Teff=20000 K, log g= 4.0. b) Teff=10000
K,  log g=4.0. c) Teff= 6000 K, log g= 4.0; here, the profiles with (thick line)
and  without (dotted line) assuming a rotation velocity of 100 km s$^{-1}$ are
shown. The profiles  are equal to those synthetized for Kurucz (thin line); the
difference at  Teff$\leq$7000 K is due to the effect of the rotation and the
contribution of other atomic transitions in the spectral wavelength ranges
synthetized in our grid.}

\figcaption{Observed spectra of stars from the Jacoby, Hunter \&
Christian (1984)  catalogue. The spectra are normalized fitting a third order
polynomial to the spectral range from 3700 to 3900 \AA\ through the continuum
windows defined  in Table 3. The spectra are compared with our synthetic spectra,
which have been  convolved to a resolution of 4 \AA\ corresponding to the
resolution of the observations. }

\figcaption{Normalized spectra of the star HD24760 (B0.5V) observed with
a dispersion of 0.4 \AA/pix compared with the synthetic spectra of T$_{eff}$= 27500
K and gravity log g=4.0.  }

\figcaption{Normalized spectra of the star HD31295 (A0V) observed with a
dispersion of 0.4 \AA/pix compared with the synthetic spectra of T$_{eff}$= 9500 K
and gravity log g=4.0.  }

 \clearpage

\begin{deluxetable}{ll} 
\footnotesize 
\tablecaption{Spectral ranges synthesized.} 
\tablewidth{0pt} 
\tablehead{ 
\colhead{Wavelength (\AA)} & \colhead{Lines} 
} 
\startdata
3720--3920 & H13, H12, H11, H10, HeI $\lambda$3819, H9, H8   \nl
3990--4150 & HeI $\lambda$4009,  HeI $\lambda$4026, H$\delta$, HeI $\lambda$4121, HeI $\lambda$4144 \nl
4300--4400 &  H$\gamma$, HeI $\lambda$4388  \nl
4420--4550 & HeI $\lambda$4471  \nl
4820--4950 &  H$\beta$, HeI $\lambda$4922\nl
\enddata 
\end{deluxetable}

\begin{deluxetable}{llllllllllllll} 
\footnotesize 
\tablecaption{Grid of models: Effective temperature and gravity covered (X).} 
\tablewidth{0pt} 
\tablehead{ 
\colhead{Teff (K)} & \colhead{5.0} & \colhead{4.5} & \colhead{4.0}& \colhead{3.5} & \colhead{3.0} & \colhead{2.5} & \colhead{2.0} & \colhead{1.5} & \colhead{1.0} & \colhead{0.5} & \colhead{0.0}  & \colhead{Model} 
} 
\startdata
50000& X & X & X &    &   &   &   &   &   &   &   & NLTE \nl
45000& X & X & X &    &   &   &   &   &   &   &   & NLTE \nl
40000& X & X & X & X &   &   &   &   &   &   &   & NLTE \nl
37500& X & X & X & X &   &   &   &   &   &   &   & NLTE \nl
35000& X & X & X & X &   &   &   &   &   &   &   & NLTE \nl
32500& X & X & X & X &   &   &   &   &   &   &   & NLTE \nl
30000& X & X & X & X &   &   &   &   &   &   &   & NLTE \nl
27500& X & X & X & X &   &   &   &   &   &   &   & NLTE \nl
25000& X & X & X & X & X &   &   &   &   &   &   & NLTE, LTE \nl
24000& X & X & X & X & X &   &   &   &   &   &   & LTE \nl
23000& X & X & X & X & X &   &   &   &   &   &   & LTE \nl
22000& X & X & X & X & X &   &   &   &   &   &   & LTE \nl
21000& X & X & X & X & X &   &   &   &   &   &   & LTE \nl
20000& X & X & X & X & X &   &   &   &   &   &   & LTE \nl
19000& X & X & X & X & X & X &   &   &   &   &   & LTE \nl
18000& X & X & X & X & X & X &   &   &   &   &   & LTE \nl
17000& X & X & X & X & X & X &   &   &   &   &   & LTE \nl
16000& X & X & X & X & X & X &   &   &   &   &   & LTE \nl
15000& X & X & X & X & X & X &   &   &   &   &   & LTE \nl
14000& X & X & X & X & X & X &   &   &   &   &   & LTE \nl
13000& X & X & X & X & X & X &   &   &   &   &   & LTE \nl
12000& X & X & X & X & X & X &   &   &   &   &   & LTE \nl
11000& X & X & X & X & X & X &   &   &   &   &   & LTE \nl
10000& X & X & X & X & X & X & X &   &   &   &   & LTE \nl
 9500 & X & X & X & X & X & X & X &   &   &   &   & LTE \nl
 9000 & X & X & X & X & X & X & X & X &   &   &   & LTE \nl
 8500 & X & X & X & X & X & X & X & X & X &   &   & LTE \nl
 8000 & X & X & X & X & X & X & X & X & X &   &   & LTE \nl
 7500 & X & X & X & X & X & X & X & X & X & X &   & LTE \nl
 7000 & X & X & X & X & X & X & X & X & X & X &   & LTE \nl
 6500 & X & X & X & X & X & X & X & X & X & X &   & LTE \nl
 6000 & X & X & X & X & X & X & X & X & X & X & X & LTE \nl
 5500 & X & X & X & X & X & X & X & X & X & X & X & LTE \nl
 5000 & X & X & X & X & X & X & X & X & X & X & X & LTE \nl
 4500 & X & X & X & X & X & X & X & X & X & X & X & LTE \nl 
 4000 & X & X & X & X & X & X & X & X & X & X & X & LTE \nl
\enddata 
\end{deluxetable}

\begin{deluxetable}{llc}
\footnotesize
\tablecaption{Line and continuum windows 
\tablenotemark{a}}
\tablewidth{0pt}
\tablehead{
\colhead{Line} & \colhead{Line window} & \colhead{Continuum window} 
} 
\startdata
 H10           & 3783--3813                       &  1 \nl
 H9             & 3823--3853                       &  1  \nl
 H8             & 3874--3904                       &  1  \nl
 H$\delta$  & 4070--4130; 4086--4116 &  2 \nl
 H$\gamma$&4311--4371; 4830--4890 & 3 \nl
 H$\beta$     & 4830--4890; 4846--4876 & 4 \nl
 HeI $\lambda$3819 & 3816--3823         & 1 \nl
 HeI $\lambda$4009 & 4006--4012         & 2 \nl
HeI $\lambda$4026  & 4020--4031         & 2 \nl
HeI $\lambda$4121  & 4118--4124         & 2 \nl
HeI $\lambda$4143  & 4139--4148         & 2 \nl
HeI $\lambda$4388  & 4381--4395         & 3 \nl
HeI $\lambda$4471  & 4464--4478         & 5 \nl
HeI $\lambda$4922  & 4915--4929         & 6 \nl
\enddata
\tablenotetext{a} {Definition of the continuum windows:
(1): 3726-3729, 3740-3743, 3760-3762,3782-3785, 3811-3812, 3869-3870, 3909-3911.
(2): 4019-4020,4037-4038, 4060-4061, 4138-4140, 4148-4150.
(3): 4301--4305, 4310-4312, 4316-4318, 4377-4381, 4392-4394, 4397-4398.
(4): 4820-4830, 4890-4900.
(5): 4439-4440, 4445-4446, 4478-4479, 4503-4506.
(6): 4900-4902, 4905-4906, 4915-4916, 4928-4929, 4948-4950.}
\end{deluxetable}

\begin{deluxetable}{rrrrrrrrr} 
\footnotesize 
\tablecaption{Equivalent width (\AA) of H$\delta$.} 
\tablewidth{0pt} 
\tablehead{ 
\colhead{Teff (K)} & \colhead{} & \colhead{} & \colhead{}& \colhead{log g} & \colhead{} & \colhead{} & \colhead{} & \colhead{} \nl
\colhead{} & \colhead{5.0} & \colhead{4.5} & \colhead{4.0}& \colhead{3.5} & \colhead{3.0} & \colhead{2.5} & \colhead{2.0} & \colhead{1.5}
} 
\startdata
50000 & 2.8   & 2.4   & 2.0   &         &         &         &  & \nl
45000 & 3.3   & 2.7   & 2.3   &         &         &         &  & \nl
40000 & 3.8   & 3.2   & 2.6   & 1.7   &         &         &  & \nl 
37500 & 4.3   & 3.6   & 2.9   & 2.1   &         &         &  & \nl
35000 & 5.3   & 4.2   & 3.2   & 2.3   &         &         &  &  \nl
32500 & 6.3   & 5.1   & 3.9   & 2.6   & 1.1   &         &  & \nl
30000 & 7.0   & 6.1   & 4.9   & 3.4   &         &         &  & \nl
27500 & 7.5   & 6.5   & 5.5   & 4.3   & 2.5   &         &  & \nl 
25000 & 8.0   & 6.8   & 5.8   & 4.8   & 3.6   &         &  &  \nl
24000 & 8.1   & 7.2   & 6.1   & 4.7   & 3.1   &         &  &  \nl
23000 & 8.4   & 7.5   & 6.2   & 5.1   & 3.4   &         &  &  \nl
22000 & 8.6   & 7.7   & 6.4   & 5.0   & 4.0   &         &  &  \nl
21000 & 8.9   & 8.0   & 6.6   & 5.3   & 4.2   &         &  &  \nl
20000 & 9.4   & 8.2   & 6.8   & 5.5   & 4.3   &         &  &  \nl
19000 & 9.7   & 8.4   & 7.0   & 5.6   & 4.4   & 3.0   &  &  \nl
18000 & 10.3 & 8.9   & 7.2   & 5.8   & 4.6   & 3.2   &  & \nl
17000 & 10.8 & 9.2   & 7.5   & 6.0   & 4.7   & 3.4   &  & \nl
16000 & 11.5 & 9.5   & 7.8   & 6.3   & 4.7   & 3.4   &  & \nl
15000 & 12.2 & 10.2 & 8.3   & 6.6   & 5.1   & 3.6   &  & \nl
14000 & 13.2 & 11.0 & 9.0   & 7.2   & 6.6   & 3.9   &  & \nl
13000 & 14.4 & 12.0 & 9.8   & 7.7   & 6.0   & 4.4   &  & \nl
12000 & 16.0 & 13.4 & 10.9 & 8.6   & 6.7   & 4.9   &  & \nl
11000 & 18.0 & 15.4 & 12.5 & 9.8   & 7.8   & 5.7   &  & \nl
10000 & 19.7 & 17.7 & 14.9 & 12.0 & 9.4   & 7.0   & 4.7   & \nl 
9500   & 19.7 & 18.7 & 16.3 & 13.5 & 10.8 & 8.0   & 5.6   &  \nl
9000   & 19.1 & 18.8 & 17.5 & 15.0 & 12.6 & 9.7   & 6.8   & 4.1 \nl
8500   & 16.9 & 17.5 & 17.5 & 16.1 & 14.6 & 11.8 & 8.8   & 5.8 \nl
8000   & 13.3 & 17.6 & 15.4 & 15.7 & 15.9 & 13.8 & 11.4 & 8.4 \nl
7500   & 10.6 & 15.1 & 12.9 & 13.0 & 14.8 & 14.7 & 13.1 & 10.9 \nl
7000   & 7.9   & 11.1 & 8.6   & 8.9   & 12.4 & 12.4 & 12.5 & 11.9 \nl
6500   & 8.1   & 8.2   & 8.3   & 8.5   & 8.7   & 8.9   & 10.4 & 10.3 \nl
6000   & 7.4   & 7.5   & 7.5   & 7.4   & 7.5   & 7.5   & 7.6   & 7.5 \nl
5500   & 7.5   & 7.4   & 7.4   & 7.4   & 7.4   & 7.4   & 7.3   & 7.3 \nl
\enddata
\end{deluxetable}

\begin{deluxetable}{rrrrrrrrr} 
\footnotesize 
\tablecaption{Equivalent width (\AA) of H$8$.} 
\tablewidth{0pt} 
\tablehead{ 
\colhead{Teff (K)} & \colhead{} & \colhead{} & \colhead{log g}& \colhead{} & \colhead{} & \colhead{} & \colhead{} & \colhead{} \nl
\colhead{} & \colhead{5.0} & \colhead{4.5} & \colhead{4.0}& \colhead{3.5} & \colhead{3.0} & \colhead{2.5} & \colhead{2.0} & \colhead{1.5}   
}
\startdata
50000 & 2.1 & 2.0 & 1.7 &  &  &  &  & \nl
45000 & 2.5 & 2.2 & 1.9 &  &  &  &  & \nl
40000 & 2.9 & 2.5 & 2.0 & 1.4 &  &  &  & \nl 
37500 & 3.2 & 2.8 & 2.2 & 1.6 &  &  &  &  \nl
35000 & 3.7 & 3.2 & 2.5 & 1.8 &  &  &  & \nl
32500 & 4.0 & 3.5 & 2.9 & 2.0 &  &  &  &  \nl
30000 & 4.6 & 3.9 & 3.1 & 2.3 &  &  &  &  \nl
27500 & 5.1 & 4.6 & 3.6 & 2.6 &  &  &  &  \nl
25000 & 5.6 & 5.0 & 4.2 & 3.2 & 1.6 &  &  & \nl
24000 & 5.9 & 5.2 & 4.1 & 3.1 & 1.9 &  &  & \nl
23000 & 6.2 & 5.4 & 4.3 & 3.2 & 2.1 &  &  & \nl
22000 & 6.3 & 5.6 & 4.6 & 3.5 & 2.3 &  &  & \nl
21000 & 6.6 & 5.9 & 4.8 & 3.7 & 2.6 &  &  & \nl
20000 & 6.8 & 6.1 & 5.1 & 4.0 & 2.8 &  &  & \nl
19000 & 7.1 & 6.3 & 5.4 & 4.2 & 3.1 & 1.7 &  & \nl 
18000 & 7.4 & 6.6 & 5.6 & 4.5 & 3.4 & 2.0 &  & \nl
17000 & 7.7 & 6.9 & 5.9 & 4.8 & 3.7 & 2.3 &  & \nl
16000 & 8.2 & 7.3 & 6.2 & 5.1 & 3.9 & 2.6 &  & \nl
15000 & 8.6 & 7.7 & 6.6 & 5.4 & 4.2 & 2.9 &  & \nl
14000 & 9.3 & 8.3 & 7.1 & 5.9 & 4.6 & 3.3 &  & \nl
13000 & 10.0 & 9.0 & 7.7 & 6.4 & 5.1 & 3.7 &  & \nl
12000 & 10.8 & 9.8 & 8.6 & 7.2 & 5.7 & 4.2 &  & \nl
11000 & 11.8 & 11.1 & 9.7 & 8.2 & 6.6 & 5.0 &  & \nl
10000 & 12.1 & 12.2 & 11.4 & 9.7 & 8.1 & 6.2 & 3.9 & \nl 
9500 & 12.0 & 12.3 & 12.0 & 11.1 & 9.2 & 7.2 & 4.9 & \nl
9000 & 11.5 & 12.1 & 12.5 & 11.9 & 10.5 & 8.7 & 6.4 & 3.9 \nl
8500 & 11.3 & 11.6 & 12.0 & 12.1 & 11.7 & 10.4 & 8.3 & 5.5 \nl
8000 & 9.5 & 10.9 & 11.3 & 11.8 & 11.8 & 11.5 & 10.1 & 8.2 \nl
7500 & 8.4 & 8.8 & 10.1 & 10.5 & 10.9 & 11.4 & 11.0 & 9.8 \nl
7000 & 7.4 & 7.5 & 7.7 & 8.0 & 9.70 & 9.7 & 10.0 & 10.3 \nl 
6500 & 6.8 & 6.8 & 6.8 & 6.3 & 7.10 & 7.3 & 8.7 & 8.8 \nl
\enddata
\end{deluxetable}

\begin{deluxetable}{lllllll} 
\footnotesize 
\tablecaption{Equivalent width (\AA) of HeI $\lambda$4471.} 
\tablewidth{0pt} 
\tablehead{ 
\colhead{Teff (K)} & \colhead{} & \colhead{} & \colhead{log g}& \colhead{} & \colhead{} & \colhead{} \nl
\colhead{} & \colhead{5.0} & \colhead{4.5} & \colhead{4.0}& \colhead{3.5} & \colhead{3.0} & \colhead{2.5}
}
\startdata
50000 & 0.36 & 0.23 & 0.11 &  &  & \nl
45000 & 0.63 & 0.45 & 0.28 &  &  &  \nl
40000 & 0.93 & 0.72 & 0.52 & 0.30 &  & \nl
37500 & 1.1 & 0.85 & 0.65 & 0.43 &  & \nl
35000 & 1.2 & 0.95 & 0.76 & 0.54 &  & \nl
32500 & 1.4 & 1.1 & 0.81 & 0.61 &  & \nl
30000 & 1.5 & 1.2 & 0.94 & 0.64 &  & \nl
27500 & 1.7 & 1.3 & 1.0 & 0.76 & 0.47 & \nl
25000 & 1.8 & 1.4 & 1.1 & 0.81 & 0.60 & \nl
24000 & 2.1 & 1.7 & 1.3 & 0.97 & 0.61 & \nl
23000 & 2.2 & 1.7 & 1.4 & 1.0 & 0.67 & \nl
22000 & 2.2 & 1.8 & 1.4 & 1.1 & 0.73 & \nl
21000 & 2.2 & 1.8 & 1.5 & 1.1 & 0.80 & \nl
20000 & 2.2 & 1.8 & 1.5 & 1.2 & 0.85 & \nl
19000 & 2.0 & 1.7 & 1.5 & 1.2 & 0.91 & 0.55 \nl
18000 & 1.8 & 1.6 & 1.4 & 1.2 & 0.93 & 0.60 \nl
17000 & 1.6 & 1.4 & 1.3 & 1.2 & 0.94 & 0.66 \nl
16000 & 1.3 & 1.2 & 0.11 & 1.0 & 0.88 & 0.67 \nl
15000 & 1.1 & 1.0 & 0.92 & 0.91 & 0.80 & 0.64 \nl
14000 & 0.81 & 0.78 & 0.74 & 0.72 & 0.68 & 0.58 \nl
13000 & 0.55 & 0.56 & 0.57 & 0.55 & 0.54 & 0.49 \nl
12000 & 0.37 & 0.40 & 0.40 & 0.41 & 0.41 & 0.39 \nl
11000 & 0.25 & 0.28 & 0.30 & 0.30 & 0.31 & 0.30 \nl
10000 & 0.19 & 0.22 & 0.25 & 0.26 & 0.26 & 0.26 \nl
\enddata
\end{deluxetable}

\begin{deluxetable}{lllllll} 
\footnotesize 
\tablecaption{Equivalent width (\AA) of HeI $\lambda$3819.} 
\tablewidth{0pt} 
\tablehead{ 
\colhead{Teff (K)} & \colhead{} & \colhead{} & \colhead{log g}& \colhead{} & \colhead{} & \colhead{} \nl
\colhead{} & \colhead{5.0} & \colhead{4.5} & \colhead{4.0}& \colhead{3.5} & \colhead{3.0} & \colhead{2.5}
} 
\startdata
50000 & 0.10 & 0.06 &         &         &         &  \nl
45000 & 0.25 & 0.19 & 0.10 &         &         &   \nl 
40000 & 0.45 & 0.39 & 0.28 & 0.10 &         &   \nl 
37500 & 0.54 & 0.50 & 0.39 & 0.22 &         &   \nl
35000 & 0.63 & 0.59 & 0.47 & 0.33 &         &   \nl
32500 & 0.70 & 0.68 & 0.56 & 0.39 &         &   \nl
30000 & 0.79 & 0.74 & 0.63 & 0.46 &         &   \nl
27500 & 0.86 & 0.83 & 0.71 & 0.51 &         &   \nl
25000 & 0.94 & 0.88 & 0.77 & 0.61 & 0.34 &  \nl 
24000 & 1.1   & 1.0   & 0.89 & 0.68 & 0.40 &  \nl
23000 & 1.1   & 1.0   & 0.94 & 0.72 & 0.44 &  \nl
22000 & 1.1   & 1.1   & 0.98 & 0.77 & 0.50 &  \nl
21000 & 1.0   & 1.1   & 1.0   & 0.80 & 0.54 &        \nl
20000 & 0.95 & 1.1   & 1.0   & 0.87 & 0.60 &        \nl
19000 & 0.87 & 1.0   & 1.0   & 0.88 & 0.64 & 0.37 \nl 
18000 & 0.71 & 0.90 & 0.96 & 0.87 & 0.67 & 0.42 \nl
17000 & 0.57 & 0.76 & 0.87 & 0.83 & 0.69 & 0.46 \nl
16000 & 0.42 & 0.63 & 0.77 & 0.76 & 0.65 & 0.47 \nl
15000 & 0.30 & 0.49 & 0.63 & 0.65 & 0.61 & 0.45 \nl
14000 & 0.21 & 0.33 & 0.49 & 0.54 & 0.53 & 0.43 \nl
13000 &         & 0.23 & 0.35 & 0.41 & 0.43 & 0.37  \nl
12000 &         &         & 0.26 & 0.31 & 0.34 & 0.29 \nl
11000 &         &         & 0.17 & 0.23 & 0.26 & 0.27 \nl
10000 &         &         &         & 0.19 &        &         \nl
\enddata
\end{deluxetable}


\clearpage

\begin{thebibliography}{}

\bibitem[]{} Auer, L. H. \& Mihalas, D. 1972, ApJS, 205, 24
\bibitem[]{} Barnard, A.J., Cooper, J., \& Smith, E.W. 1974, J.Q.S.R.T., 14, 1025
\bibitem[]{} Bianchi, L., Hutchings, J.B., \& Massey, P. 1996, A\&A, 111, 2303
\bibitem[]{} Burstein, D., Faber, S.M., Gaskell, C.M., \& Krumm, N. 1984, ApJ, 287, 586
\bibitem[]{} Cananzi, K., Augarde, R. \& Lequeux, J. 1993, A\&AS, 101, 599
\bibitem[]{} Gabler, R., Gabler, A., Kudritzki, R.P., \& Pauldrach, A. 1989, A\&A, 226, 162 
\bibitem[]{} Herrero, A., Kudritzki, R.P., V\'\i lchez, J.M., Butler, K., \& Haser, S. 1992, A\&A, 261, 209
\bibitem[]{} Hubeny, I. 1988, Compt. Phys. Com., 52, 103
\bibitem[]{} Hubeny, I., Harmanec, P. \& Stefl, S. 1986, Bull. Astron. Inst. Czechosl. 37, 370
\bibitem[]{} Hubeny, I. \& Lanz, T. 1995a, ApJ, 439, 875
\bibitem[]{} -. 1995b, TLUSTY-A User's Guide
\bibitem[]{} Hubeny, I.,  Lanz, T.,\& Jeffery, C.S. 1995, SYNSPEC-A User's Guide
\bibitem[]{} Jacoby, G.H., Hunter, D. A., \& Christian, C. A. 1984, ApJS, 56, 257
\bibitem[]{} Jones, J. A. 1997, Ph.D. Thesis, Univ. of North Carolina, Chapel Hill
\bibitem[]{} Kurucz, R. L. 1979, ApJS, 40, 1
\bibitem[]{} -. 1993, CD-ROM 13, ATLAS9 Stellar Atmosphere 
Programs and 2 km/s Grid (Cambridge: Smithsoniam Astrophys. Obs.)
\bibitem[]{} Kurucz, R.L. \& Peytremann, E. 1975, SAO Spec. Rep. No. 362
\bibitem[]{} Shamey, L. 1969, PhD Thesis, University of Colorado
\bibitem[]{} Vidal, C. R., Cooper, J. \& Smith, E. W. 1973, ApJS, 25, 37
\bibitem[]{} Walborn, N. R. \& Fitzpatrick, E. L. 1990, PASP, 102, 379
\end{thebibliography}
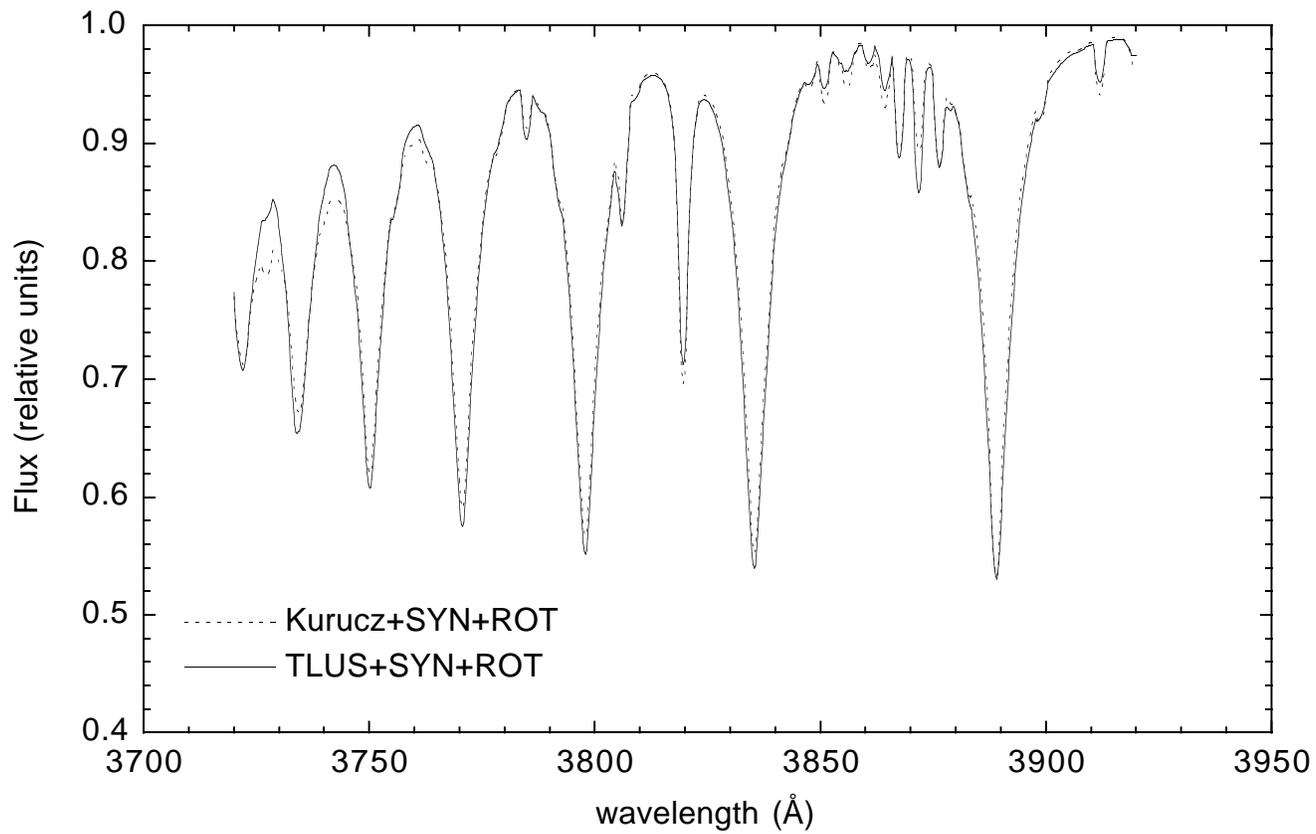
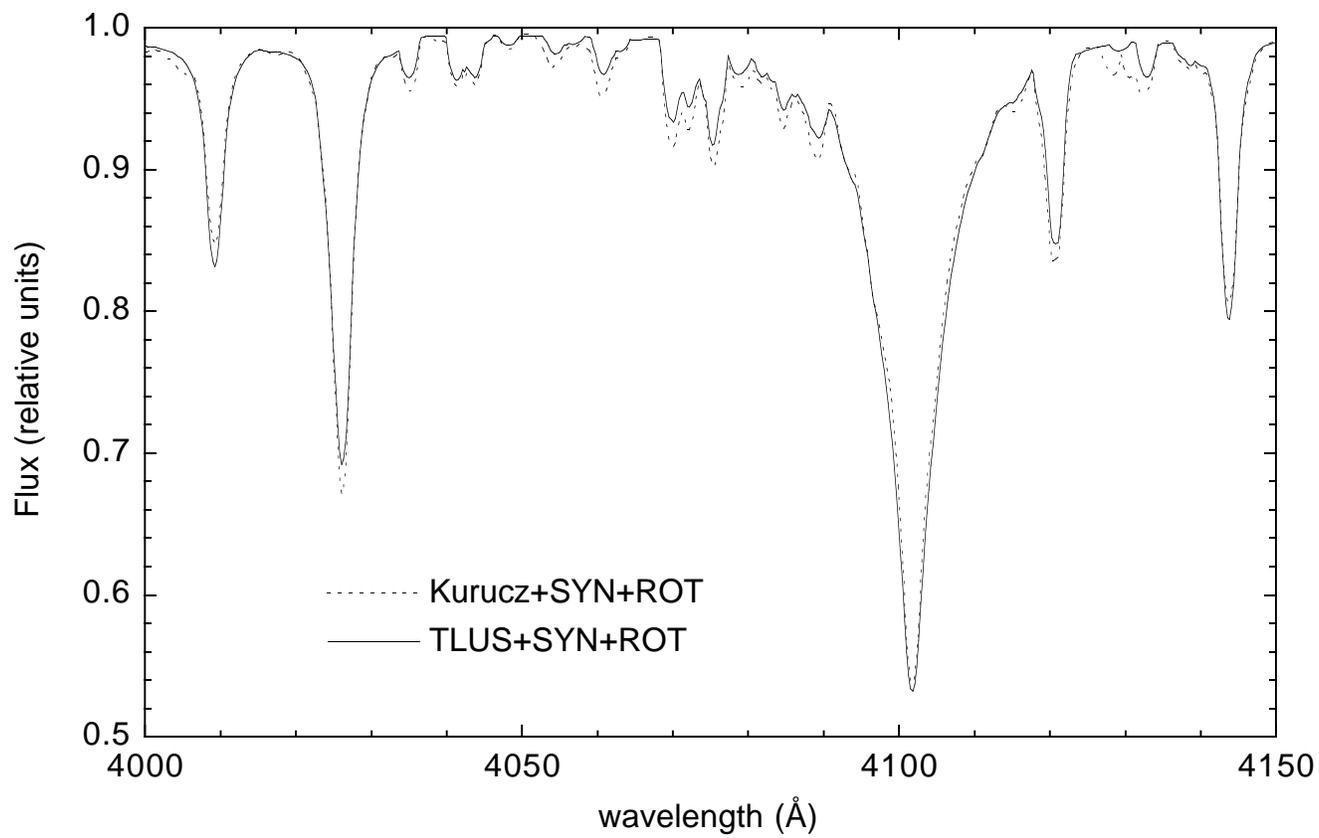
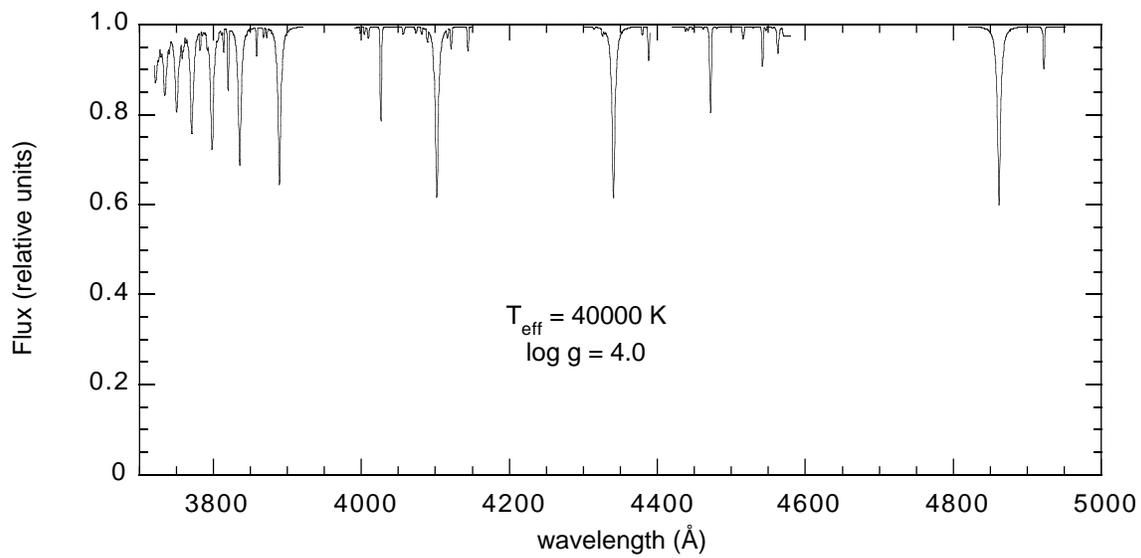
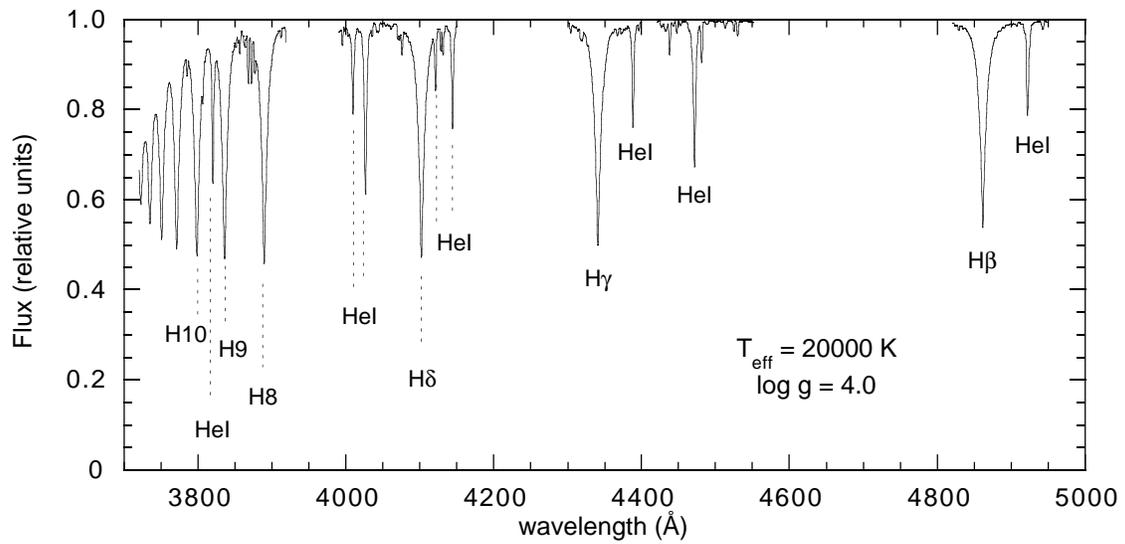
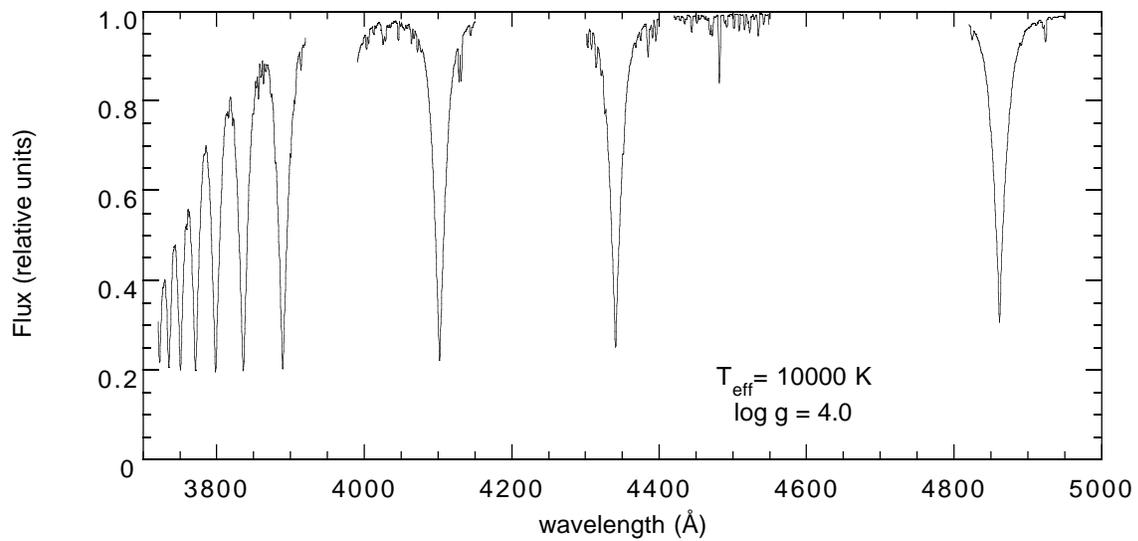
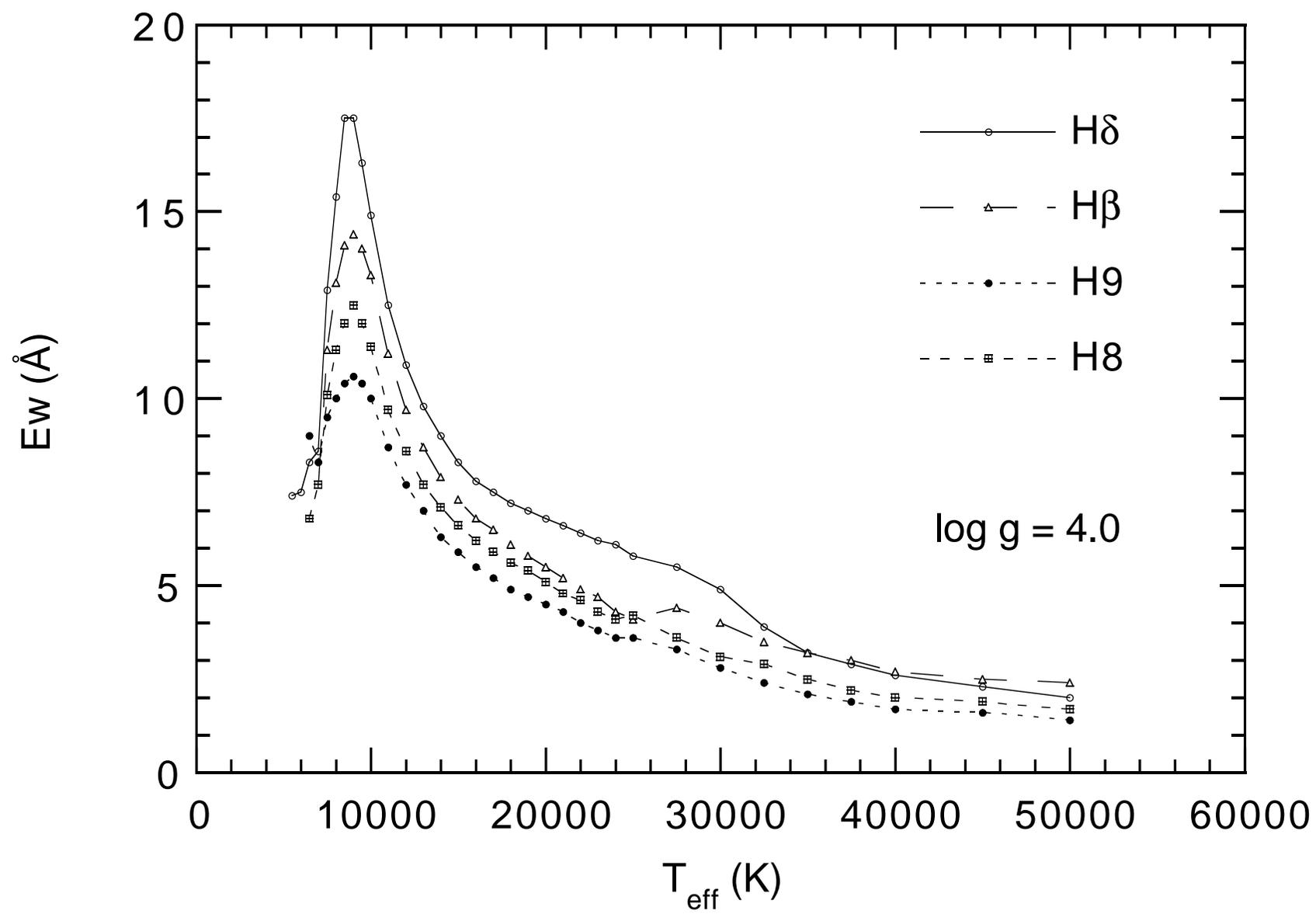
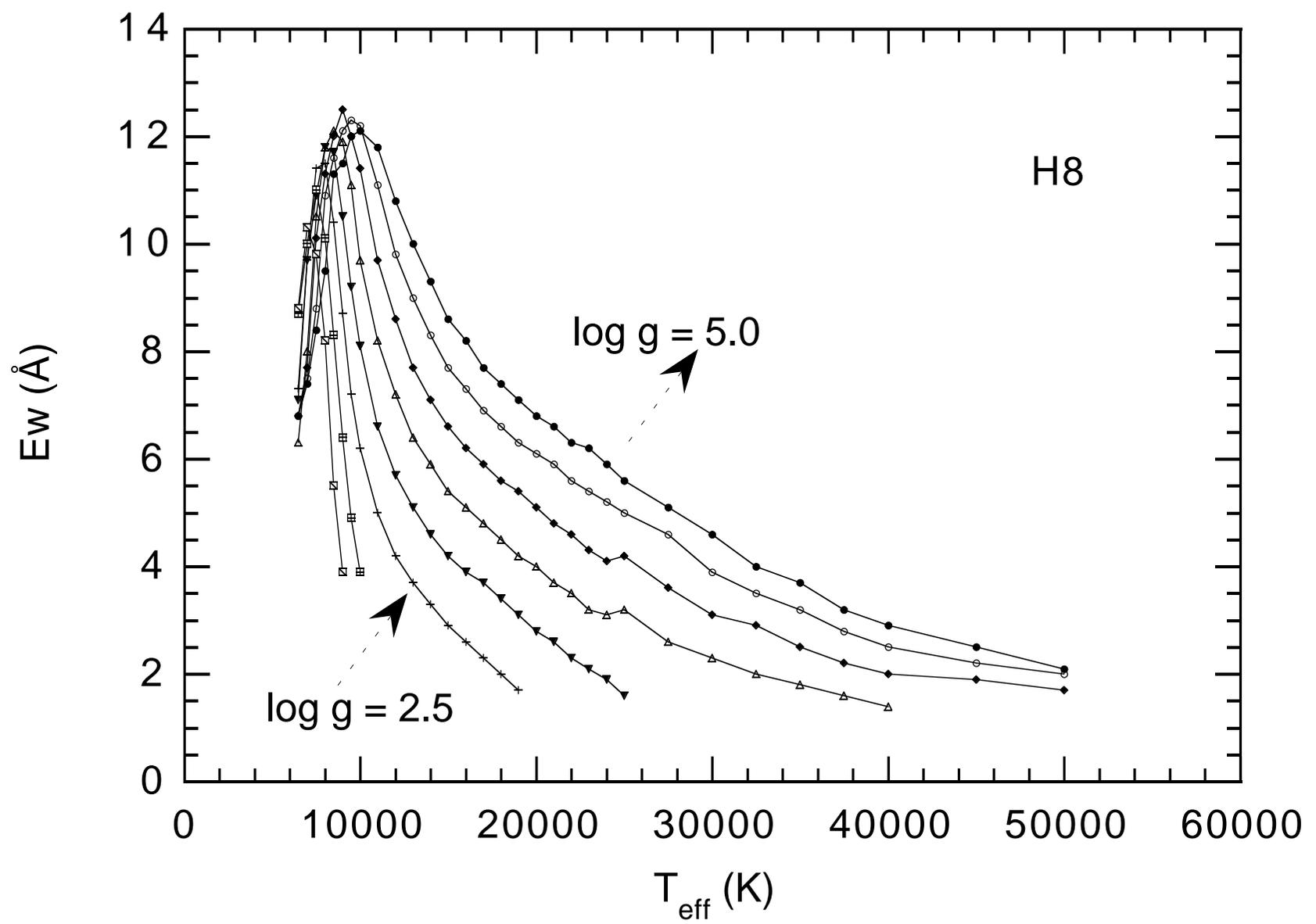
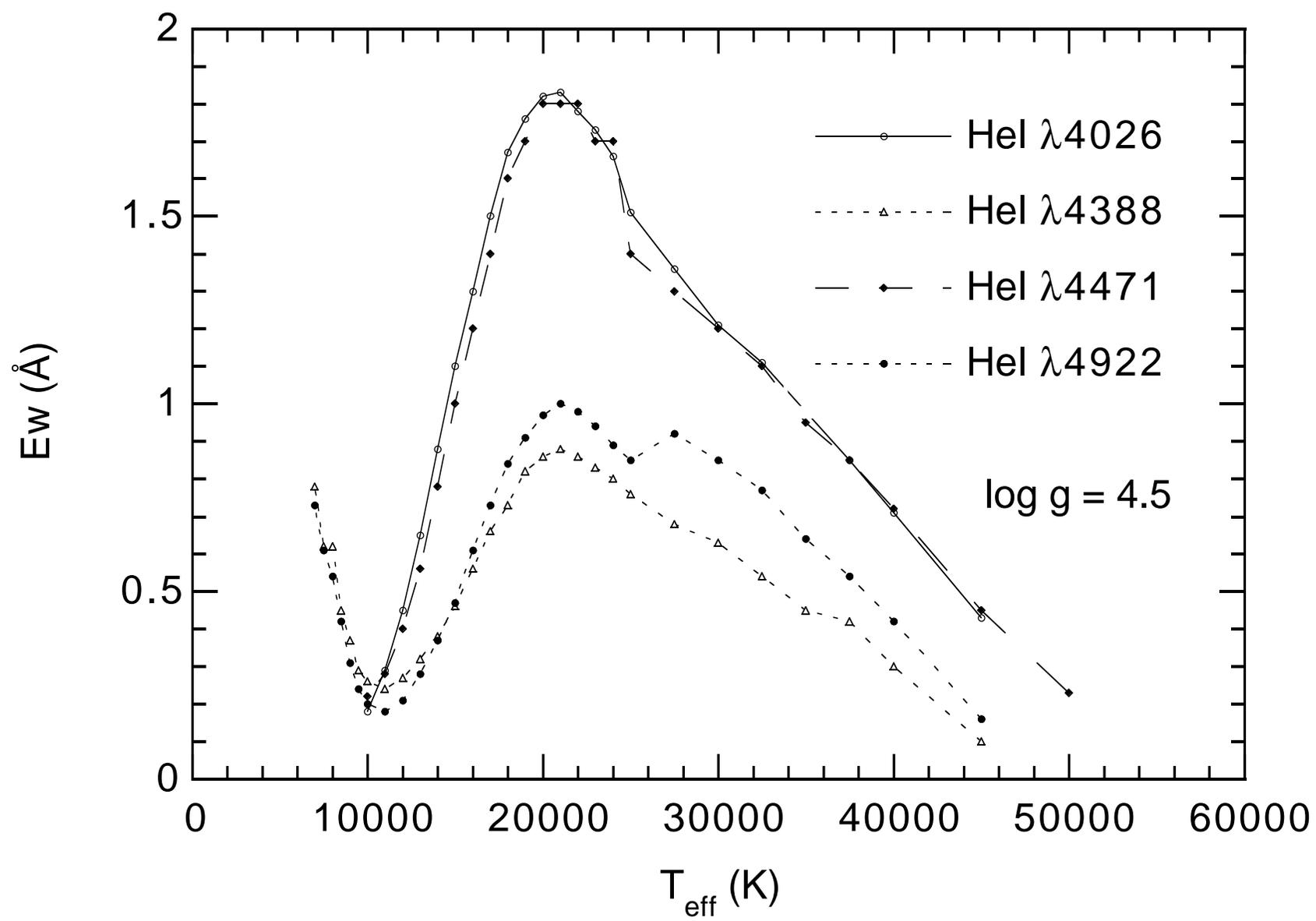
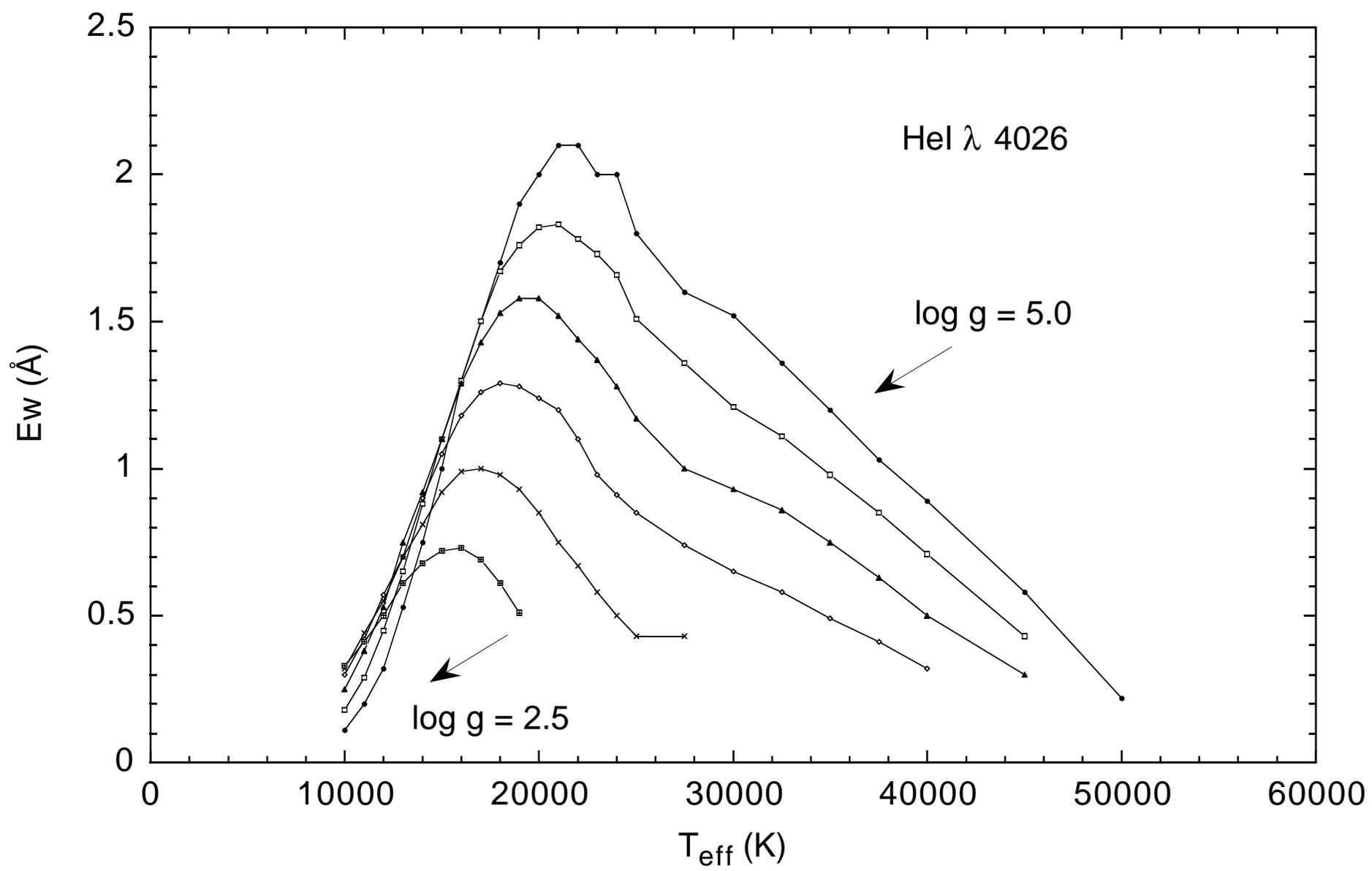
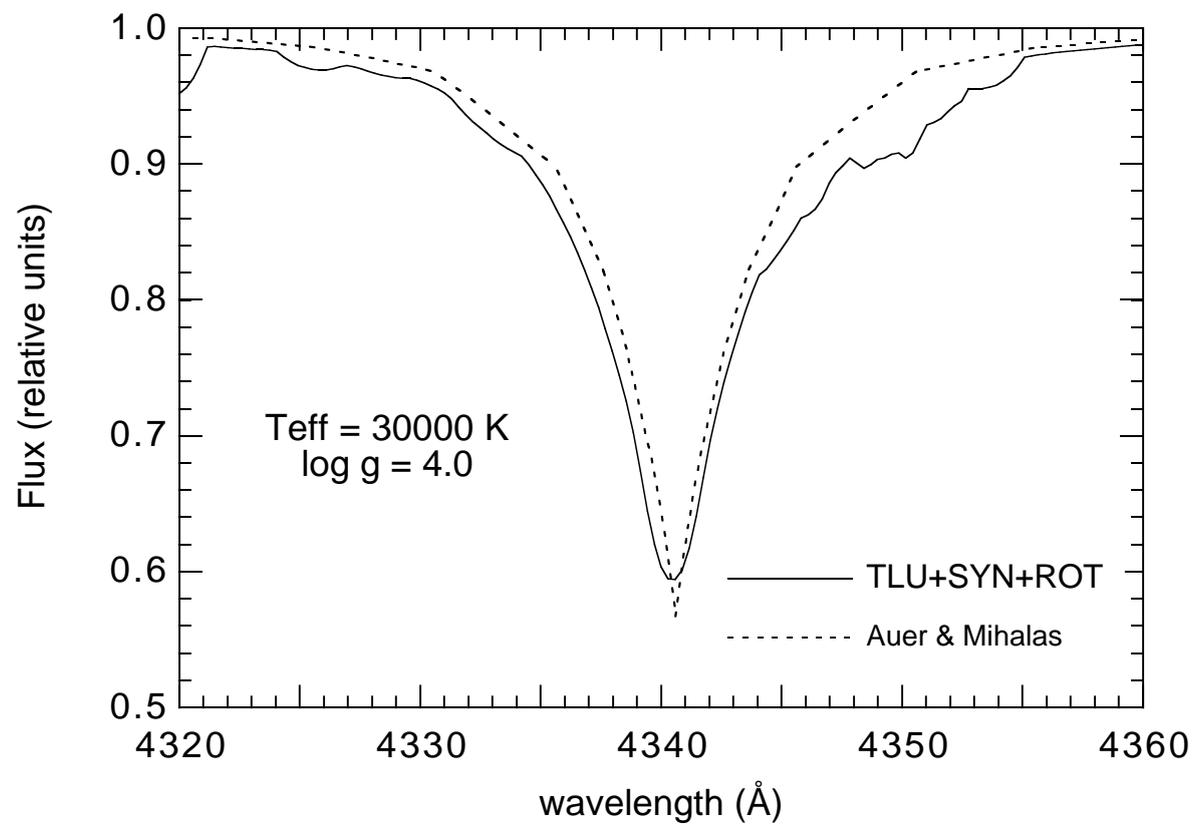
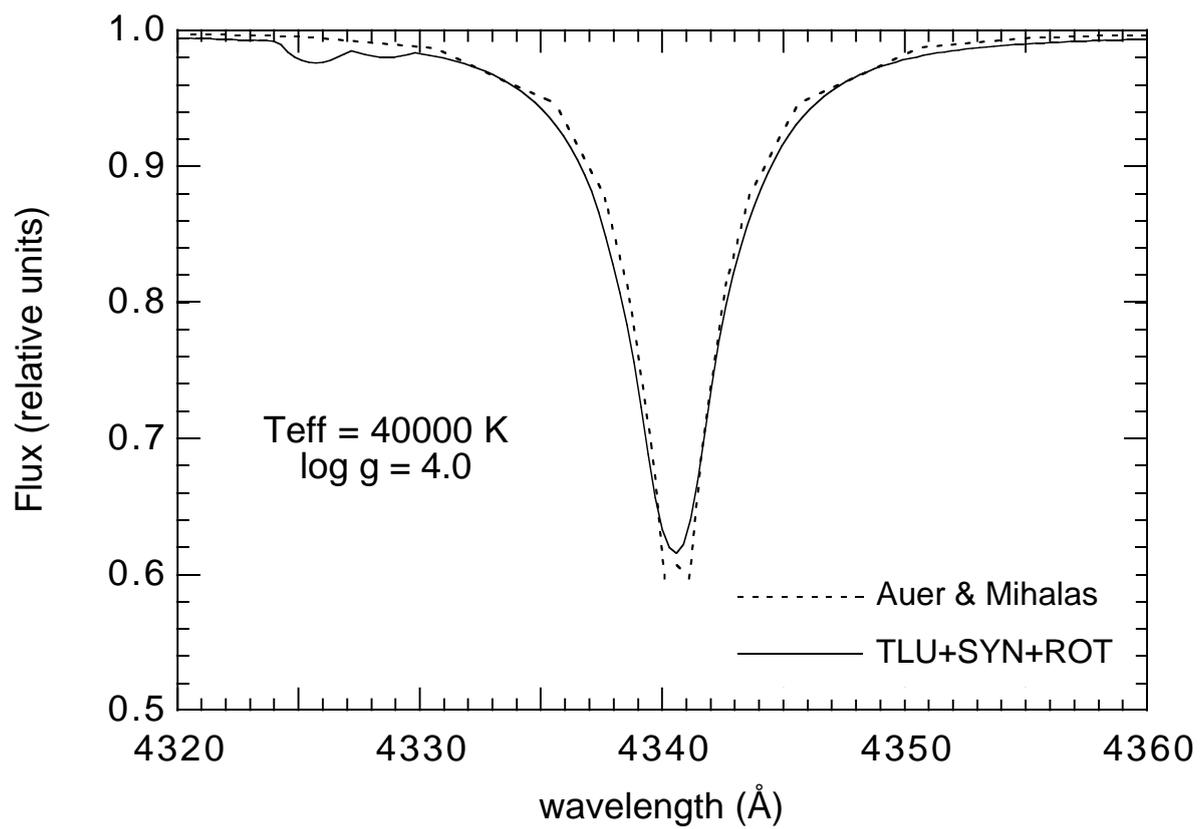
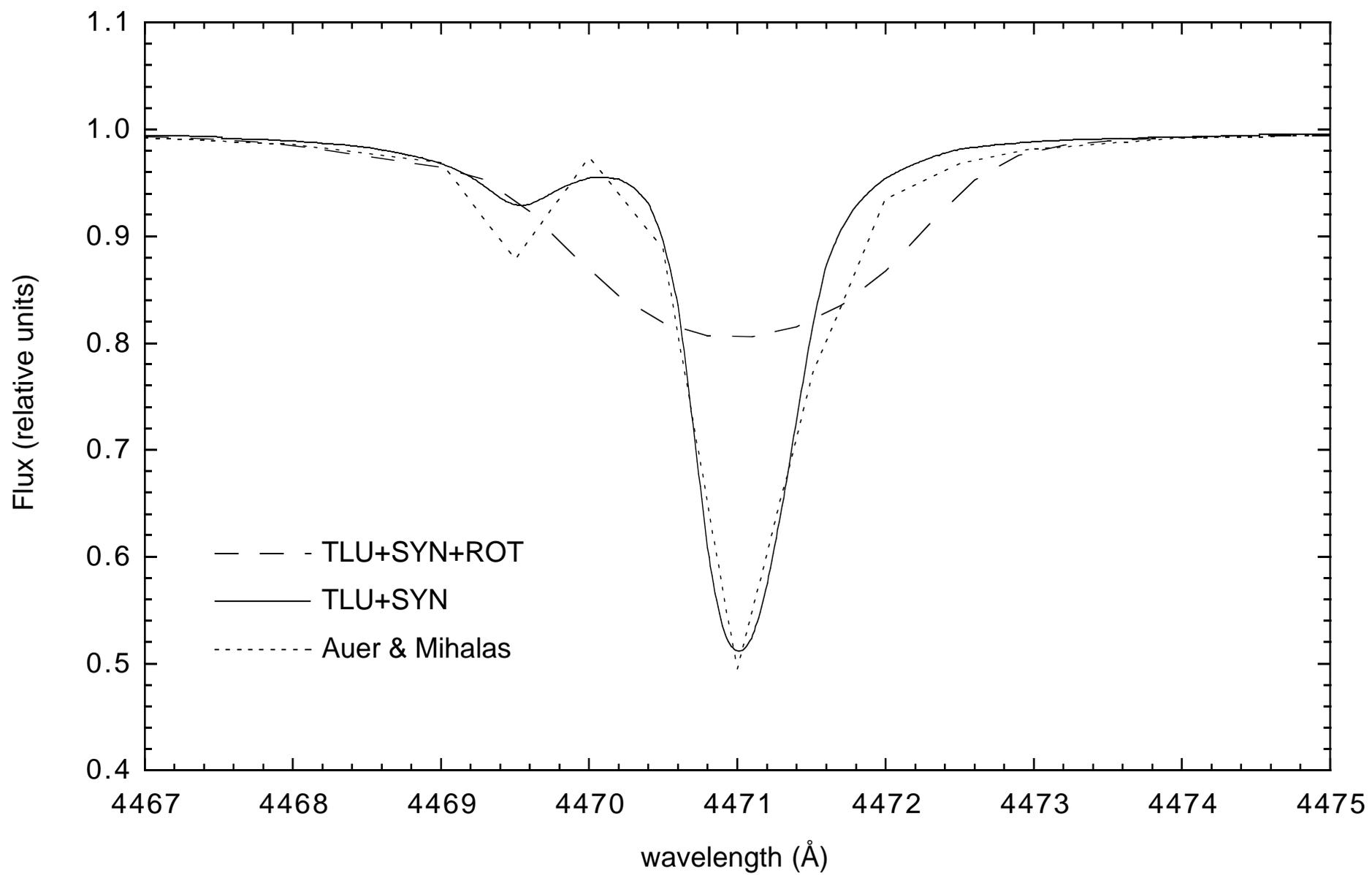
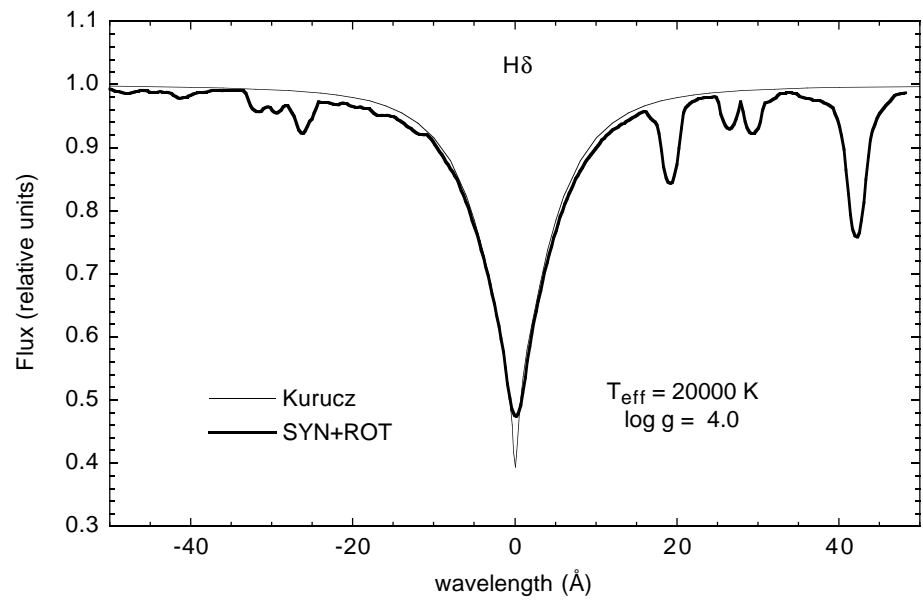
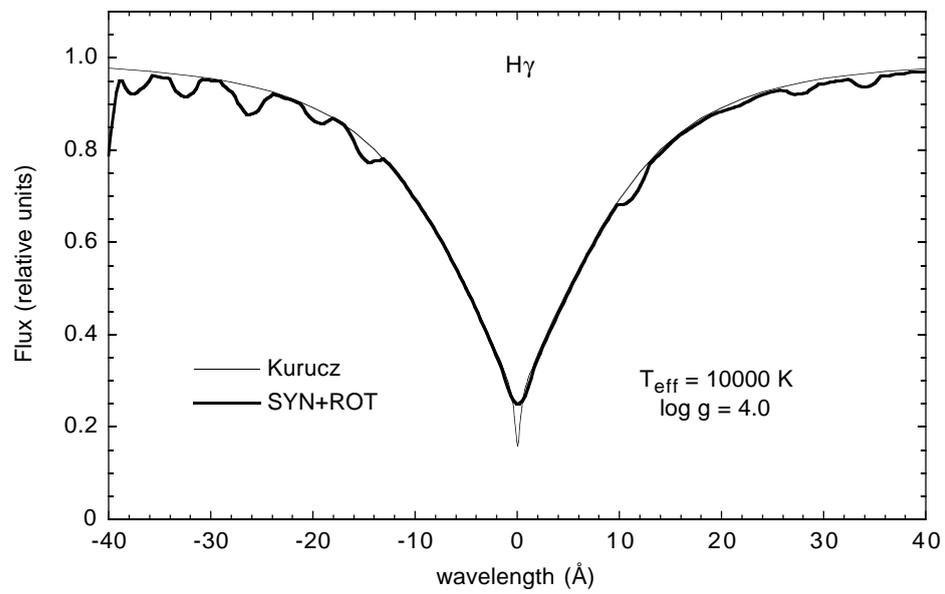
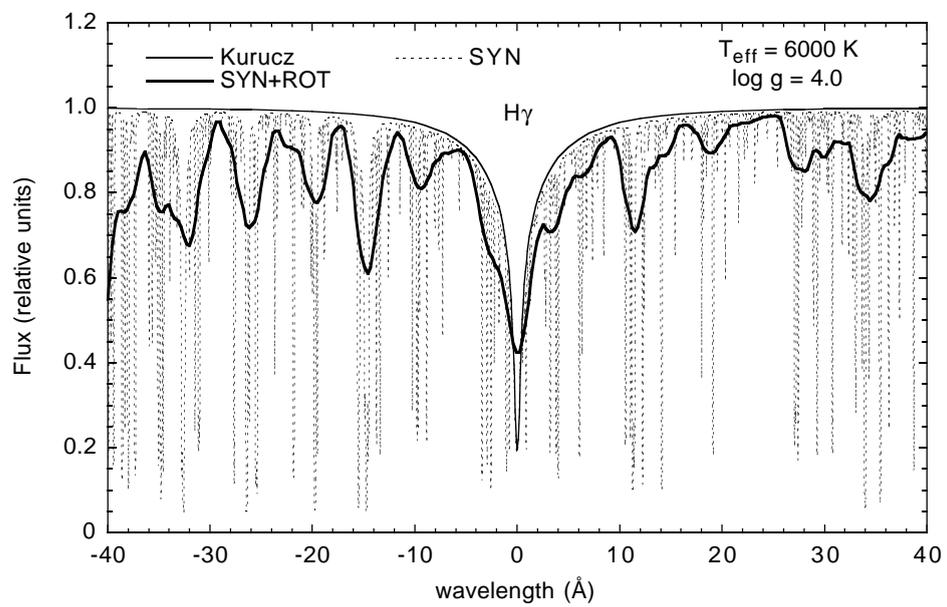
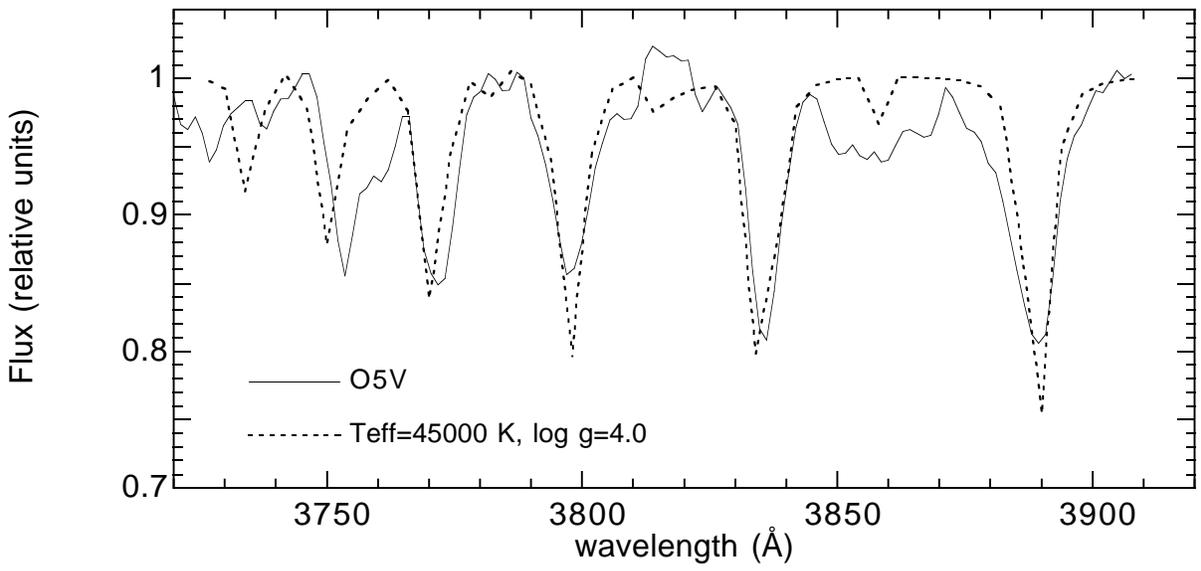
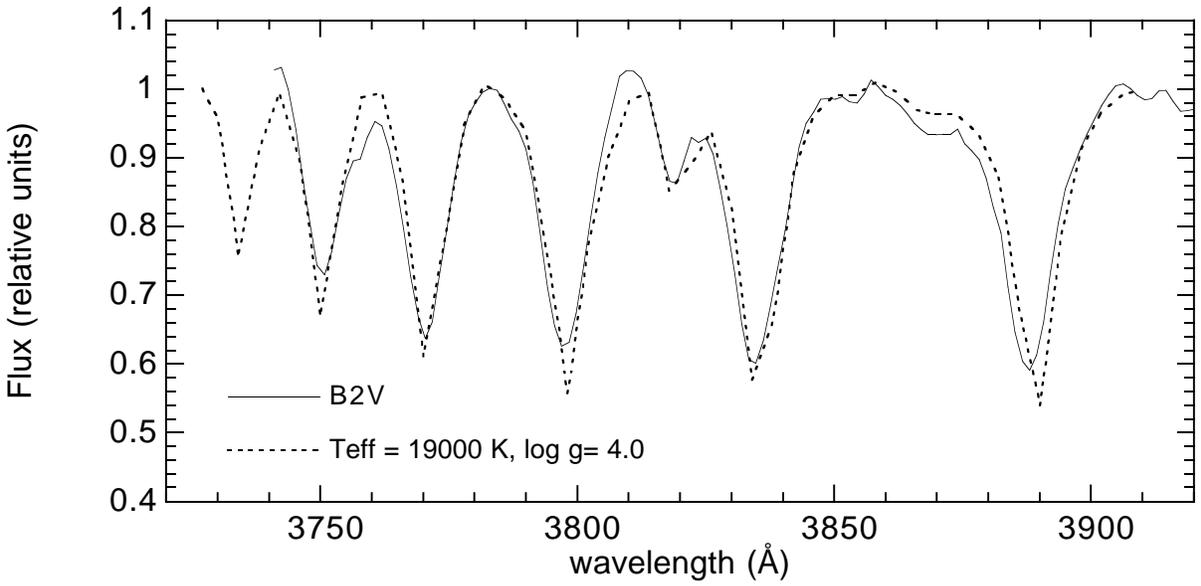
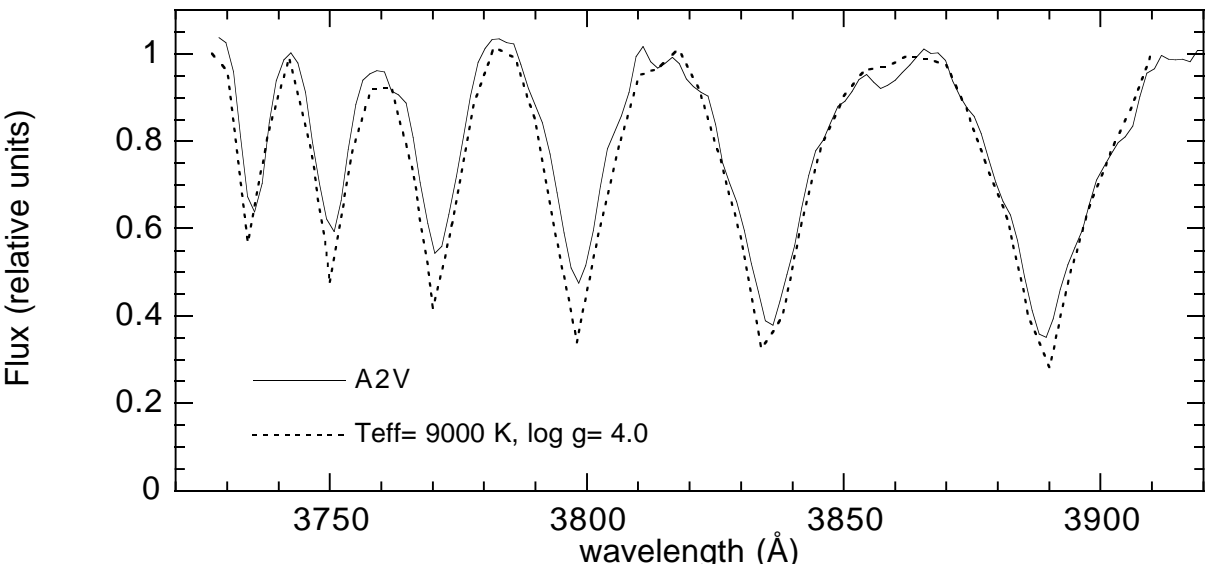
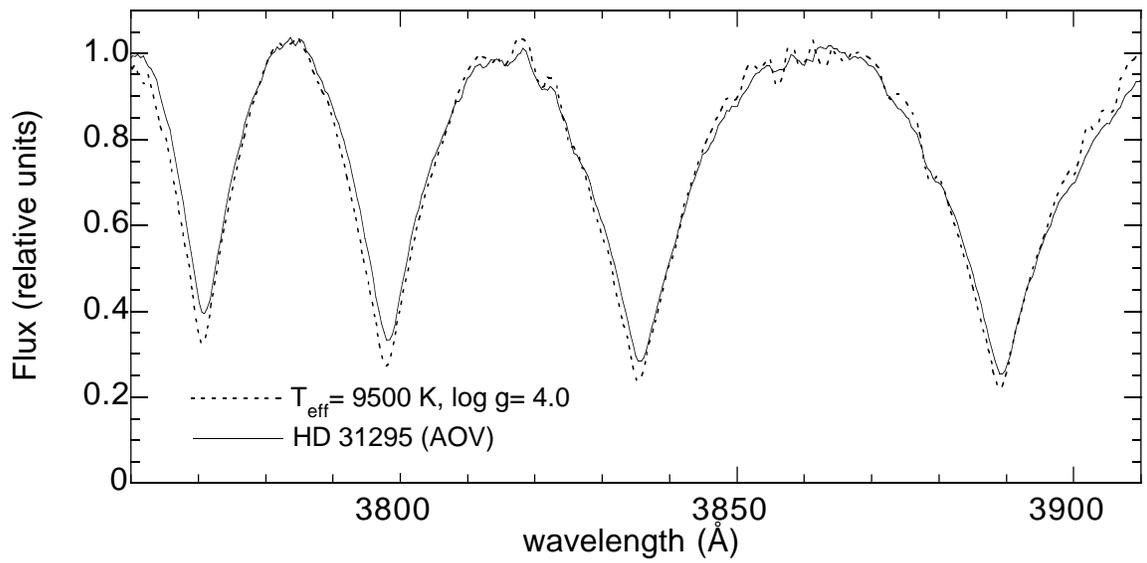
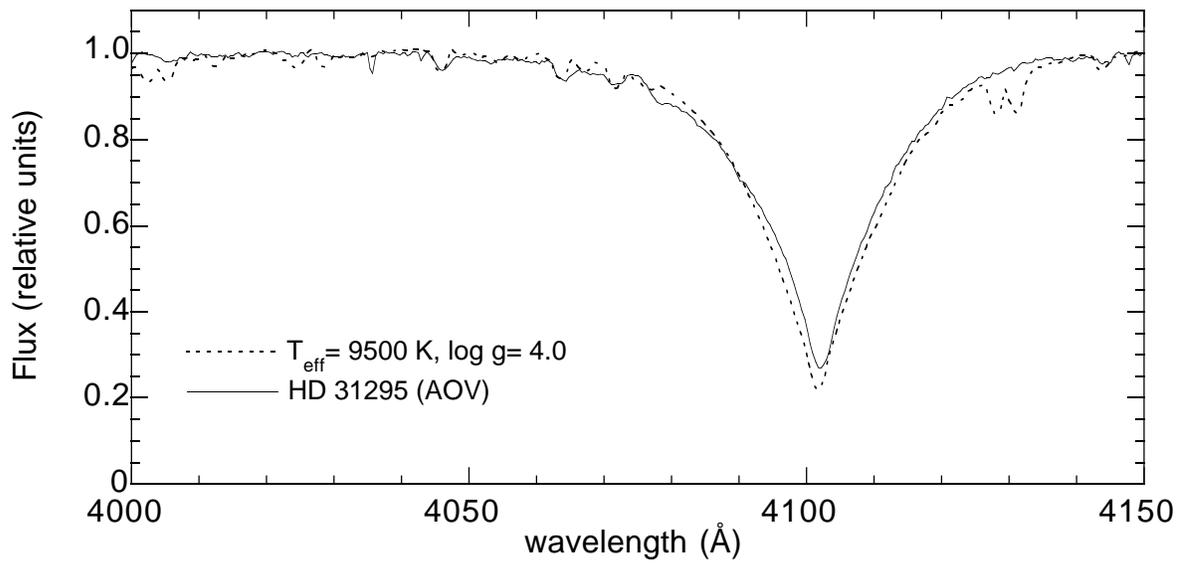
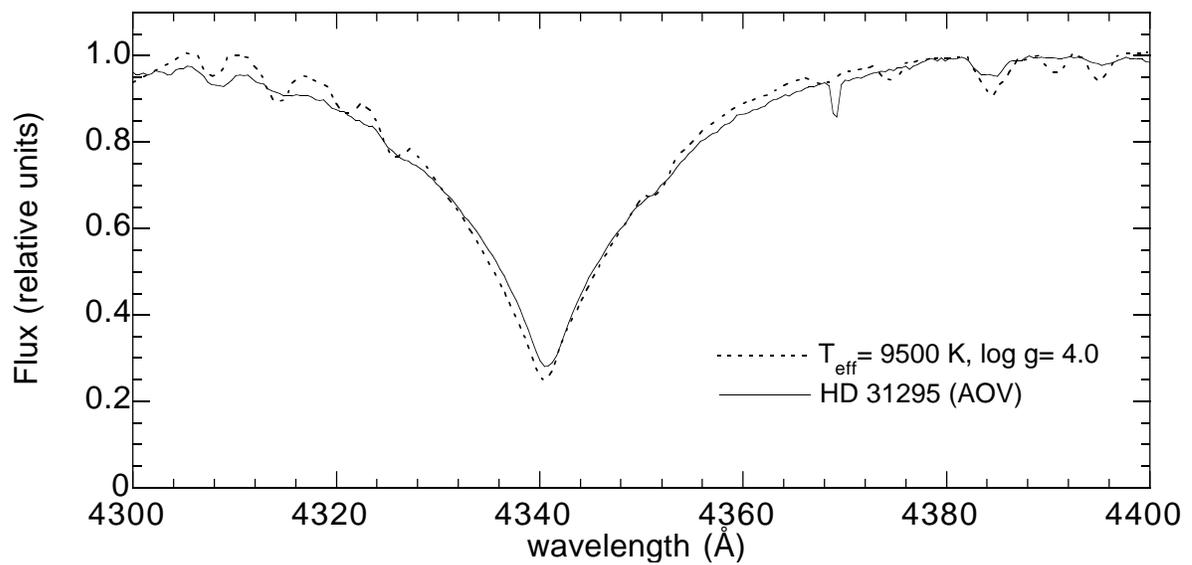
\end{document}